\documentclass[prd,onecolumn,english]{revtex4}
\usepackage{babel}

\usepackage{amsmath}
\usepackage{amssymb}
\usepackage{amsthm} 

\usepackage{array}

\usepackage{slashed}

\usepackage[dvips]{graphicx} 
\usepackage{subfigure}

\usepackage{varioref}   
\usepackage{xr}         

\usepackage{pstricks}           

\allowdisplaybreaks[1]

\newcommand\nn{\nonumber}
\newcommand\ba{\begin{eqnarray}}
\newcommand\ea{\end{eqnarray}}
\newcommand\eq[1] {\begin{align} #1 \end{align}}   
\newcommand\ga[1] {\begin{gather} #1 \end{gather}}   

\newcommand{\br}[1]{\left( #1 \right)}
\newcommand{\brs}[1]{\left[ #1 \right]}

\newcommand{\GeV}{\mbox{GeV}}

\newcommand{\M} {{\cal M}} 

\newcommand{\Moller} {{M{\o}ller}~} 

\begin{document}

\title{Two-loop electroweak vertex corrections for polarized \Moller scattering}
\author{A.G.~Aleksejevs}
\affiliation{Memorial University, Corner Brook, Canada}
\email{aaleksejevs@grenfell.mun.ca}

\author{S.G.~Barkanova}
\affiliation{Acadia University, Wolfville, Canada}
\email{svetlana.barkanova@acadiau.ca}

\author{Yu.M.~Bystritskiy}
\affiliation{Joint Institute for Nuclear Research, Dubna, Russia}
\email{bystr@theor.jinr.ru}

\author{\framebox{E.A.~Kuraev}}
\affiliation{Joint Institute for Nuclear Research, Dubna, Russia}
\email{kuraev@theor.jinr.ru}

\author{V.A.~Zykunov}
\affiliation{Belarusian State University of Transport, Gomel, Belarus}
\email{vladimir.zykunov@cern.ch}

\date{\today}

\begin{abstract}
The contributions to the electron-electron-photon vertex of two-loop electroweak corrections are calculated.
The relative correction to the parity-violating asymmetry of M{\o}ller scattering for the
case of $11~\GeV$ electron scattered off the electron at rest is found to be about $-0.0034$
and should be taken into account at future experiment MOLLER at JLab.
\end{abstract}

\maketitle

\section{Introduction}
\label{SectionIntroduction}

Nowadays high energy physics faces difficulties.
The energies reached at existing facilities are almost at the limit of their technical possibilities
and reasonable cost of the projects.
Besides whole set of experimental information is in a confident agreement
with Standard Model (SM) predictions, there are large number of indications that
at this scale there is a possibility for "new physics" to manifest itself.
This is mostly the content of scientific programs for existing and future
accelerators. But most probable option is that within accessible energy range
the traces of new physics can be discovered as a small deviations from
SM predictions. Revelation of new physical phenomena is only possible by comparing
of detailed experimental result with model predictions.
The aim of the present paper is to continue the elaboration of precise
description of one of the most prominent process --- M{\o}ller scattering \cite{Moeller:1932}.

This process has wide and active interest from both experimental and theoretical sides for several reasons.
It has allowed the high-precision determination of the electron-beam
polarization at SLC \cite{Swartz:1994yv}, SLAC \cite{Steiner:1998gf,Band:1997ee}, JLab \cite{Hauger:1999iv} and MIT-Bates \cite{Arrington:1992xu}
(and as a future prospects --- the ILC \cite{Alexander:2000bu} and CLIC \cite{Linssen:2012hp}).
The polarized  M{\o}ller scattering can be an excellent tool in measuring parity-violating
weak interaction asymmetries \cite{Derman:1979zc}.
The first observation of Parity Violation (PV) in the M{\o}ller scattering was made
by the E-158 experiment at SLAC \cite{Kumar:1995ym,Kumar:2007zze,Anthony:2003ub},
which studied scattering of 45- to 48-GeV polarized electrons on the
unpolarized electrons of a hydrogen target.
It results at $Q^2 = -t = 0.026~\GeV^2$ for the observable parity-violating asymmetry
$A_{PV} = (1.31 \pm 0.14\ \mbox{(stat.)} \pm 0.10\ \mbox{(syst.)}) \times 10^{-7}$ \cite{Anthony:2005pm} which allowed one
of the most important SM parameters --- the sine of the  Weinberg angle $\sin \theta_W$ --- to be determined
with the best accuracy at that moment.

The MOLLER (Measurement Of a Lepton Lepton Electroweak Reaction) experiment planned
at the Jefferson Lab aims to measure the parity-violating asymmetry in the scattering of
$11~\GeV$ longitudinally-polarized electrons from the atomic electrons in a liquid
hydrogen target with a combined stati\-sti\-cal and systematic
uncertainty of 2\,\%~\cite{vanOers:2010zz,Benesch:2011,Kumar:2009zzk,Benesch:2014bas}.
With such precision any inconsistency with the SM predictions will clearly
signal the new physics. However, a comprehensive analysis of radiative corrections
is needed before any conclusions can be made. Since MOLLER's stated precision
goal is significantly more ambitious than that of its predecessor E-158,
{\it theoretical input} for this measurement must include not only a full treatment of one-loop
(next-to-leading order, NLO) electroweak radiative corrections (EWC) but also
two-loop corrections (next-to-next-leading order, NNLO).
Although, two-loop corrections to the cross section may seem to be small,
it is much harder to estimate their scale and behavior for such a complicated observable as
the parity-violating asymmetry to be measured by the MOLLER experiment.

This paper is the part of our attempt to perform this {\it theoretical input}.
The significant efforts have been already dedicated to one-loop EWC.
A short review of the references on that topic is given in \cite{Aleksejevs:2010ub,Aleksejevs:2012zz},
where we calculated a full set of the one-loop EWC
both numerically with no simplifications using computer algebra packages
and by-hand in a compact form analytically free from nonphysical parameters.
One way to find some indication
of the size of higher-order (two-loop) contributions is to compare results that are expressed in
terms of quantities related to different renormalization schemes.
In \cite{Aleksejevs:2010nf} we provided a tuned comparison between the result obtained with different
renormalization conditions, first within one scheme then between two schemes.
Our calculations in the on-shell and Constrained Differential Renormalization schemes show the difference of about 11\%,
which is comparable with the difference of 10\% between
$\rm \overline{MS}$ \cite{Czarnecki:1995fw} and the on-shell scheme  \cite{Petriello:2002wk}.

The two-loop EWC to the Born cross section ($\sim \M_0\M_0^+$) can be divided onto two classes:
$Q$-part induced by quadratic one-loop amplitudes $ \sim \M_1\M_1^+$,
and $T$-part -- the interference of Born and two-loop amplitudes $ \sim 2 \, \mbox{\rm Re} \br{\M_{0} \M_{2}^+}$
(here index $i$ in the amplitude $\M_i$ corresponds to the order of perturbation theory).
The $Q$-part was calculated exactly in \cite{Aleksejevs:2011de}
(using Feynman--t'Hooft gauge and the on-shell renormalization),
where we show that the $Q$-part is much higher than the planned experimental uncertainty of MOLLER,
i.e. the two-loop EWC is larger than was assumed in the past.
The large size of the $Q$-part demands detailed and consistent treatment of $T$-part, but this
formidable task will require several stages. Our first step was to
calculate the gauge-invariant  double boxes \cite{Aleksejevs:2012sq}.
Next step was to calculate the two-loop gauge invariant set of boson self energies and
vertices function diagrams \cite{Aleksejevs:2013gxa}.
In this paper we took into account two photon emission mechanism in
soft photon approximation.
It is important to calculate
hard photons bremsstrahlung contribution in accordance with MOLLER detector
parameters. This work was partially done for one-loop EWC in \cite{Zykunov:2015cea}.
Then we consider in \cite{Aleksejevs:2015dba} the EWC arising from the contribution
of a wide class of the gauge-invariant Feynman amplitudes of the box type with one-loop insertions:
fermion mass operators [or Fermion Self-Energies in Boxes],
vertex functions [or Vertices in Boxes],
and polarization of vacuum for bosons [or Boson Self-Energies in Boxes].
Also, in this paper one can find extended literature review and all necessary details and notations
useful for understanding of present paper.
So, finally, this paper logically related to manuscript \cite{Aleksejevs:2015dba},
where we do the next step --- we calculate the insertions of two-loop vertices to vertices (VV),
fermion self-energies to vertices (FSEV) and double vertices (DV).

This theoretical program must be completed by testing our two-loop results with the results of computer algebra packages
(like FeynArts, FormCalc etc). This testing was already done for one-loop calculations,
to do this for two loops results is the important task for our group to be done in the next future.

The paper is organized as follows.
In Section~\ref{sec.BasicNotations} we consider the asymmetry in Born approximation and introduce the basic notations.
In Section~\ref{sec.ExtraWZ} we calculate two Feynman diagrams with extra $W$ and $Z$ boson subgraphs (VV).
Section~\ref{sec.MassOperators} is devoted to the diagrams with lepton mass operators insertions (FSEV).
In Section~\ref{sec.ComplexVertices} complex vertices are considered (DV).
And in Section~\ref{sec.NumericalResults} we give numerical estimation of total effect of these contributions.

\section{Basic notations}
\label{sec.BasicNotations}

We consider the process of electron-electron elastic scattering, i.e. \Moller process:
\eq{
    e(p_1,\lambda_1)+e(p_2,\lambda_2) \to e(p_1',\lambda_1')+ e(p_2',\lambda_2'),
}
where $\lambda_{1,2}$ ($\lambda'_{1,2}$) are the chiral states of initial (final) electrons and $p_{1,2}$ are 4-momenta of initial electrons
and $p_{1,2}'$ are 4-momenta of final electrons.
The first measurement of parity-violating (left-right) asymmetry
\eq{
A =\frac{d\sigma^{----}-d\sigma^{++++}}{d\sigma^{----}+d\sigma^{++++}+d\sigma^{+-+-}+d\sigma^{+--+}+d\sigma^{-+-+}+d\sigma^{-++-}}=
\frac{|M^{----}|^2-|M^{++++}|^2}{\sum_\lambda |M^{\lambda}|^2}
}
in M{\o}ller scattering was made
by E-158 experiment at SLAC \cite{Kumar:1995ym,Kumar:2007zze,Anthony:2003ub}.
In lowest order of perturbation theory in frames of QED the matrix element squared
which is summed over polarization states of electrons has the following form:
\eq{
\sum_\lambda |M^{\lambda}|^2=8(4\pi\alpha)^2\frac{s^4+t^4+u^4}{t^2u^2}.
}
We use the notation for the kinematic invariants neglecting of electron mass $m$:
\eq{
s=2p_1p_2, \qquad t=-2p_1p_1', \qquad u=-2p_1p_2', \qquad s+t+u=0.
}
Thus here and further we neglect the terms of order $O(m^2/s)$ since in MOLLER experiment it is expected that beam energy is
$E_{\rm beam}=11~\GeV$, that is $s = 2 m E_{\rm beam} \approx 0.01124~\GeV^2$.
Within the Standard Model one has additional contribution in Born approximation with $Z$-boson exchange which gives rise to polarization asymmetry $A^{0}$:
\eq{
A^{0}=\frac{s}{2m_W^2}A_{(0)}\frac{a}{s_W},
\qquad
A_{(0)}=\frac{y(1-y)}{1+y^4+(1-y)^4},
\qquad
y=\frac{-t}{s}=\frac{1-c}{2},
}
where $c=\cos\theta$ is the cosine of scattering angle $\theta=\br{\widehat{\vec{p}_1,\vec{p}_1^{\;'}}}$ in the system of center-of-mass of electrons,
$m_W$ is the $W$-boson mass and $a$ is the so-called "weak electron charge"
\eq{
a=1-4s_W^2.
}
Now let's recall that
$s_{W}\  (c_{W})$
is the sine (cosine) of the Weinberg angle expressed in terms of the $Z$- and $W$-boson
masses according to the Standard Model rules:
\eq{
s_{W}=\sqrt{1-c_{W}^2},\qquad
c_{W}=m_{W}/m_{Z}.
}
Thus, the factor $a$ is just
$a \approx 0.109$ and the asymmetry is therefore suppressed by both $s/m_W^2$ and $a$. Even at
Central Region (CR) of MOLLER (at $\theta \sim 90^\circ$, i.e. $t \approx u \approx -s/2$), where the Born asymmetry is maximal, this asymmetry is extremely small:
\eq{
    A^{0}=\frac{s}{9m_W^2}\frac{a}{s_W^2} \approx 9.4968 \cdot 10^{-8}.
}

\begin{figure}
    \centering
    \mbox{
        \subfigure[]{\includegraphics[width=0.15\textwidth]{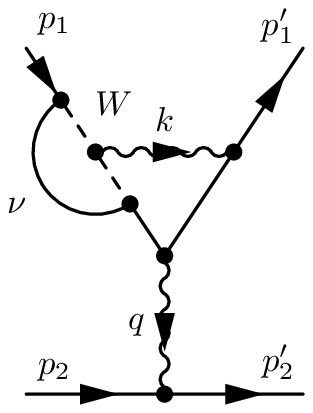}\label{fig.a}}
        \qquad
        \subfigure[]{\includegraphics[width=0.15\textwidth]{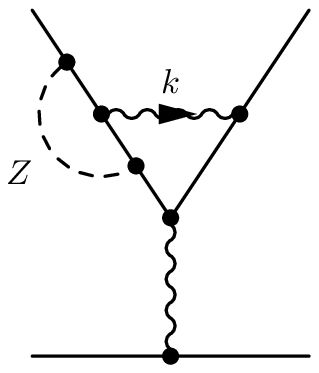}\label{fig.b}}
        \qquad
        \subfigure[]{\includegraphics[width=0.20\textwidth]{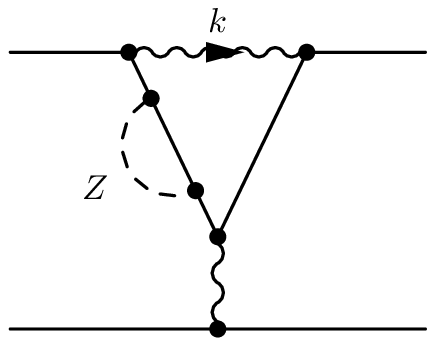}\label{fig.Z}}
        \qquad
        \subfigure[]{\includegraphics[width=0.20\textwidth]{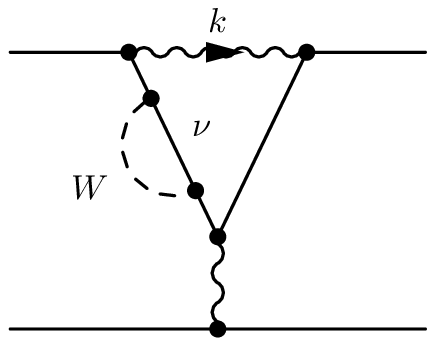}\label{fig.W}}
    }
    \nn\\
    \vspace{0.5cm}
    \mbox{
        \subfigure[]{\includegraphics[width=0.20\textwidth]{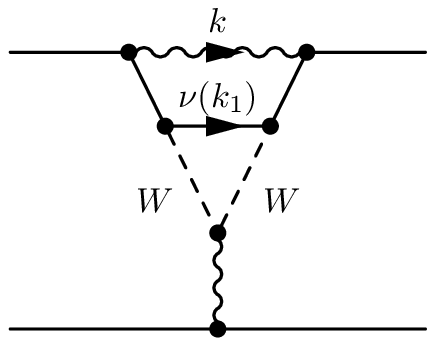}\label{fig.e}}
        \qquad
        \subfigure[]{\includegraphics[width=0.20\textwidth]{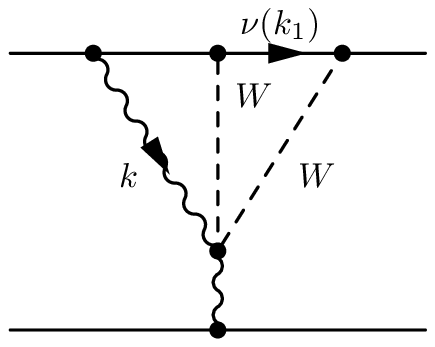}\label{fig.f}}
        \qquad
        \subfigure[]{\includegraphics[width=0.20\textwidth]{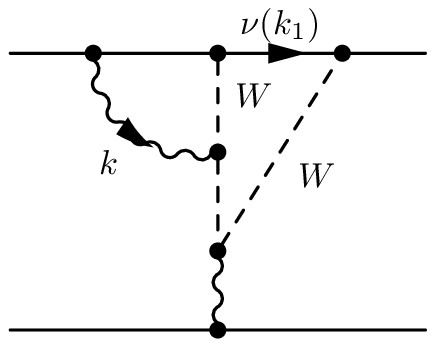}\label{fig.g}}
        \qquad
        \subfigure[]{\includegraphics[width=0.20\textwidth]{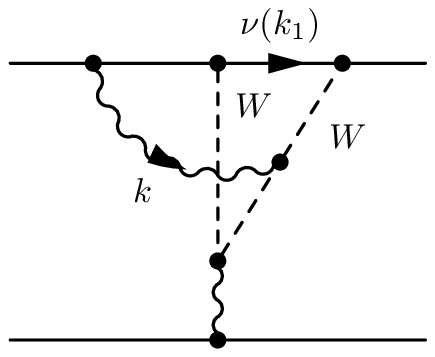}\label{fig.h}}
    }
    \caption{Two-loop vertices to vertices (a, b), fermion self-energies to vertices (c,d), double vertices (e-h).
    Photons are denoted by wavy lines, massive bosons -- by dashed lines, electrons and neutrinos -- by solid lines.
    Notations on diagrams show the type of particle ($Z, W, \nu$) and 4-momenta of them.
}
    \label{fig.TwoLoopsElectronVertex}
\end{figure}

It is the main aim of this paper to estimate the contribution of some classes
of two-loop contributions, which have some logarithmical enhancement.
As for the non-enhanced ones -- they have an order of $(-t/m_Z^2)(\alpha/\pi)^2 \approx 10^{-11}$ for the
CR of MOLLER.
Below we consider contribution to the vertex function $\Delta V_\mu$ for on-mass-shell electrons
$p_1^2=p_1'^2=m^2$ and the space-like 4-momentum of the virtual photon $Q^2=-q^2=-(p_1-p_1')^2 \gg m^2$
in two-loop level from the class of Feynman diagrams containing the intermediate states with $W$- and $Z$-bosons
(see Fig.~\ref{fig.TwoLoopsElectronVertex}).
Due to vertex renormalization condition $\Delta V_\mu|_{Q^2=0}=0$ the corresponding contribution
is proportional to $Q^2/m_{W,Z}^2$.
Thus we restrict ourselves by the condition
\eq{
    \rho_{i}^{-1} =  Q^2/m^2_i \ll 1,\ \ i=W,Z.
}

\section{Vertex subgraphs with extra $W$ and $Z$ bosons}
\label{sec.ExtraWZ}

The one-loop level expressions for $WW\gamma$-vertex (see subgraph of process $e(p_1)\to e(p_1-k)+\gamma(k)$ where electron with momentum $(p_1-k)$ is off-mass-shell in Fig.~\ref{fig.a}) has a form:
\eq{
V_\mu^{a}(p_1,k)&=-ie\bar{u}(p_1-k)\gamma_\mu\omega_- u(p_1) \frac{g^2}{32\pi^2}I^{a}(k^2), \label{eq.VW}
\\
I^{a}(k^2)&=\int\limits_0^1 d y \int\limits_0^y d x\br{6\ln\frac{m_W^2 y-k^2x\bar{x}}{m_W^2 y}+\frac{k^2 x(1-2x)}{m_W^2 y-k^2x\bar{x}}}, \label{eq.IW}
}
where $\omega_\pm =1 \pm \gamma_5$, $g=e/s_W$ and $e$ is the electron charge value ($e=|e|>0$). Here and below we use the common notation $\bar x \equiv 1-x$, $\bar y \equiv 1-y$, etc.
Analogously, one-loop vertex with one additional $Z$-boson (see corresponding subgraph in Fig.~\ref{fig.b}) looks like:
\eq{
V_\mu^{b}(p_1,k)&=-ie\bar{u}(p_1-k)\gamma_\mu(a\pm\gamma_5)^2 u(p_1)\frac{g^2}{(4c_W)^2 16\pi^2} I^{b}(k^2),
\\
I^{b}(k^2)&=\int\limits_0^1 d y \int\limits_0^y d x\br{-2\ln\frac{m_Z^2 \bar{y}-k^2x\bar{x}}{m_Z^2 \bar{y}}+\frac{k^2 x\bar{x}}{m_Z^2 \bar{y}-k^2x\bar{x}}}.
}
Both functions $I^{a,b}(k^2)$ are normalized as $I^{a,b}(0)=0$.

Integrating over 4-momenta of photon $k$ we get the contribution of (\ref{eq.VW}) into full two-loop vertex $V_\mu$ (see Fig.~\ref{fig.a}) as
\eq{
\Delta V^a_\mu&=-ie\frac{g^24\pi\alpha}{(16\pi^2)^2}I^{a}_\mu, \qquad
I^{a}_\mu=I_{1\mu}^{a} + I_{2\mu}^{a} = \int\frac{d^4 k}{i\pi^2}\frac{ N^a_\mu}{k^2(k^2-2p_1k)(k^2-2p_1'k)}I^{a}(k^2), \label{eq.ImuW} \\
N^a_\mu&=\bar{u}(p_1')\gamma_\lambda(\hat{p}_1'-\hat{k}+m)\gamma_\mu(\hat{p}_1-\hat{k}+m)\gamma^\lambda\omega_- u(p_1),
}
where two terms $I_{1,2\mu}^{a}$ in (\ref{eq.ImuW}) correspond to two terms in (\ref{eq.IW}).
%
Omitting the terms of order ${\cal O}(\rho_i^{-2})$ we write down the contribution $I_{1\mu}^{a}$ as
\eq{
I^{a}_{1\mu}=6\int\limits_0^1 y\bar{y} d y\int\frac{d^4 k}{i\pi^2}\left.\frac{N^a_\mu}{m_W^2(k^2-2p_1k)(k^2-2p_1'k)}\right|_{|k^2| \ll m_W^2}.
}
Using Feynman parameters trick one can integrate over loop momenta $k$ and obtain the result as a sum
of ultra-violet finite (UV-finite) and ultra-violet divergent (UVD) parts:
\eq{
\frac{Q^2}{m_W^2}\int\limits_0^1  x\bar{x}\br{\ln\frac{m_W^2}{b^2}-1}d x + \text{UVD-part},\qquad
b^2=(p_1x+p_1'\bar{x})^2=m^2+Q^2x\bar{x}.
}
Here and below we use the same notations for unrenormalized and renormalized quantities.
Thus after renormalization of $I^{a}_{1\mu}$ we obtain expression
\eq{
I^{a}_{1\mu}=\frac{Q^2}{m_W^2}c_1^a\bar{u}(p_1')\gamma_\mu\omega_- u(p_1),\qquad
c_1^a=\frac{1}{3}\ln\frac{m_W^2}{Q^2}+\frac{7}{18} = 5.0406.
}
Here and everywhere below the number value corresponds to CR of MOLLER.

Second term $I_{2\mu}^{a}$ in (\ref{eq.ImuW}) can be written as
\eq{
I^{a}_{2\mu}=\int\limits_0^1 y\bar{y} d y\int\limits_0^yd x \frac{1-2x}{1-x}\int\frac{d^4 k}{i\pi^2}\frac{N^a_\mu}{(k^2-\sigma^2)(k^2-2p_1k)(k^2-2p_1'k)},
}
where $\sigma^2=m_W^2y/(x\bar{x})$. Again using standard manipulations we arrive to
\eq{
I^{a}_{2\mu}&=\bar{u}(p_1')\gamma_\mu\omega_- u(p_1)\int\limits_0^1 y\bar{y} d y\int\limits_0^yd x \frac{1-2x}{1-x}
\int\limits_0^1 d x_1\int\limits_0^1 2 y_1 d y_1 \brs{\ln\frac{\Lambda^2}{D_a}-\frac{3}{2}-\frac{Q^2}{D_a}(1-y_1x_1)(1-y_1\bar{x}_1)},
}
where $D_a=y_1^2b_a^2+\sigma^2 \bar{y}_1$, $b_a^2=\br{x_1 p_1 + \bar{x}_1 p_1'}^2 = m^2 + x_1\bar{x}_1 Q^2$ and $\Lambda$ is the UV-regularization parameter.
Applying the renormalization procedure we obtain
\eq{
I^{a}_{2\mu}&=\frac{Q^2}{m_W^2}c_2^a\bar{u}(p_1')\gamma_\mu\omega_- u(p_1), \nn \\
c_2^a&=-\int\limits_0^1 d x_1\int\limits_0^1 2 y_1 d y_1\int\limits_0^1 y \bar{y} d y\int\limits_0^y d x\frac{1-2x}{1-x}\times\nn\\
&\times\brs{\rho_W\ln\br{1+\frac{1}{\rho_W}\frac{x_1\bar{x}_1y_1^2x\bar{x}}{y\bar{y}_1}}+
\frac{\rho_W x\bar{x}(1-y_1x_1)(1-y_1\bar{x}_1)}{\rho_W y\bar{y}_1+y_1^2x\bar{x}x_1\bar{x}_1}} = -0.0930.
}
The final expression for contribution of $W$-vertex subgraph to the vertex function (see Fig.~\ref{fig.a}) is
\eq{
\Delta V^a_\mu=-ie\frac{Q^2}{m_W^2}\frac{g^24\pi\alpha}{(16\pi^2)^2}(c_1^a+c_2^a)\bar{u}(p_1')\gamma_\mu\omega_- u(p_1).
}

Contribution of $Z$-vertex subgraph (see Fig.~\ref{fig.b}) has a form
\eq{
\Delta V^b_\mu&=-ie\frac{g^2 8\pi\alpha}{(4c_W)^2(16\pi^2)^2}I^{b}_\mu, \qquad
I^{b}_\mu =\int\frac{d^4 k}{i\pi^2}\frac{N^b_\mu}{k^2(k^2-2p_1k)(k^2-2p_1'k)}I^{b}(k^2), \nn \\
N^b_\mu&=\bar{u}(p_1')\gamma_\lambda(\hat{p}_1'-\hat{k}+m)\gamma_\mu(\hat{p}_1-\hat{k}+m)\gamma_\lambda(a\pm \gamma_5)^2 u(p_1).
}
In the similar way we obtain for contribution of $Z$-vertex subgraph to the vertex function
\eq{
\Delta V^b_\mu=-ie\frac{Q^2}{m_Z^2}\frac{g^2 8\pi\alpha (1\pm a)^2}{(4c_W)^2(16\pi^2)^2}(c_1^b+c_2^b)\bar{u}(p_1')\gamma_\mu u(p_1),
}
where
\eq{
c_1^b&=\frac{2}{9}\ln\frac{m_Z^2}{Q^2}+\frac{7}{27} = 3.4164, \nn \\
c_2^b&=-\int\limits_0^1 d x_1\int\limits_0^1 2 y_1 d y_1\int\limits_0^1 y\bar{y} d y\int\limits_0^y d x\frac{1-2x}{1-x}\times\nn\\
&\times\brs{\rho_Z\ln\br{1+\frac{1}{\rho_Z}\frac{x_1\bar{x}_1y_1^2x\bar{x}}{y\bar{y}_1}}+
\frac{\rho_Z x\bar{x}(1-y_1x_1)(1-y_1\bar{x}_1)}{\rho_Z y\bar{y}_1+y_1^2x\bar{x}x_1\bar{x}_1}}
=-0.0944.
}

\section{Electroweak electron mass operator insertion to the vertex function}
\label{sec.MassOperators}

Now let's consider the set of Feynman diagrams of vertex type containing electron Mass Operator (MO)
with internal $Z$ or $W$ bosons insertions (see Fig.~\ref{fig.Z},\ref{fig.W}).
The relevant contribution to the vertex function has a form
\eq{
\Delta V^{\rm MO}_\mu&=-ie\frac{\alpha}{2\pi}\brs{V^Z_\mu+V^W_\mu}, \qquad
V^i_\mu=\int\frac{d^4 k}{i\pi^2}\frac{N^i_\mu}{k^2(k^2-2p_1'k)},
}
where $i=Z,W$ and the numerator $N^i_\mu$ is:
\eq{
N^i_\mu&=\bar{u}(p_1')\gamma_\lambda(\hat{p}_1'-\hat{k}+m)\gamma_\mu(\hat{p}_1-\hat{k}+m)\gamma^\lambda c^{(i)} u(p_1) M^i(k, p_1),
}
with
\ga{
c^{(Z)}=\frac{2g^2(a\pm \gamma_5)^2}{(4c_W)^2 8\pi^2},\qquad
c^{(W)}=\frac{g^2\omega_-}{16\pi^2}.
}
Mass operator $M^i(k, p_1)$ of electron with both external legs off-mass-shell looks like:
\eq{
    M^i(k, p_1)=\int\limits_0^1 x_1\bar{x}_1 d x_1\int\limits_0^1\frac{d z}{m_i^2-x_1z(k^2-2p_1k)}.
}
Standard Feynman procedure of joining the denominators and the loop momentum integrating gives us
\eq{
V^i_{\mu}&=\int\limits_0^1\bar{x}_1 d x_1\int\limits_0^1 \frac{d z}{z}\int\frac{d^4 k}{i\pi^2}\frac{N_\mu}{k^2(k^2-2p_1'k)(k^2-2p_1k-\sigma_i^2)},
\nn \\
N_\mu&=\bar{u}(p_1')\gamma_\lambda(\hat{p}_1'-\hat{k}+m)\gamma_\mu(\hat{p}_1-\hat{k}+m)\gamma_\lambda u(p_1),\qquad
\sigma_i^2=\frac{m_i^2}{x_1 z},
}
which can be simplified to the form:
\eq{
V^i_{\mu}=\bar{u}(p_1')\gamma_\mu u(p_1)V^i, \qquad
V^i=\int\limits_0^1\bar{x}_1 d x_1\int\limits_0^1 \frac{d z}{z}\int\limits_0^1 d x\int\limits_0^1 2y d y \br{\ln\frac{\Lambda^2}{D_i}-\frac{Q^2}{D_i}}, \qquad
D_i=b^2y^2+xy\sigma_i^2.
}
After renormalization and expansion on powers of $Q^2/m_i^2$ we obtain
\eq{
V_i=-\frac{Q^2}{m_i^2}\int\limits_0^1\bar{x}_1 d x_1\int\limits_0^1 \frac{d z}{z}\int\limits_0^1 d x\int\limits_0^1 2y d y
\brs{x_1zy\bar{x}+\frac{m_i^2}{D_i}(\bar{y}-y^2x\bar{x})}.
}
Final expression for the contribution to the vertex function (see Fig.~\ref{fig.Z},\ref{fig.W}) is
\eq{
\Delta V^{\rm MO}_\mu=ie\frac{\alpha}{2\pi}\brs{\frac{Q^2}{m_Z^2}c_3^Z\frac{g^2}{(4c_W)^2 4\pi^2}(1\pm a)^2 \bar{u}(p_1')\gamma_\mu u(p_1)
+
\frac{Q^2}{m_W^2}c_3^W\frac{g^2}{16\pi^2}\bar{u}(p_1')\gamma_\mu\omega_- u(p_1)},
}
with
\eq{
c_3^Z=\frac{1}{6}\ln\frac{m_Z^2}{Q^2}+\frac{2}{3} = 3.0345, \qquad
c_3^W=\frac{1}{6}\ln\frac{m_W^2}{Q^2}+\frac{2}{3} = 2.9925.
}

\section{Contribution of diagrams containing $WW\gamma,\ WW\gamma\gamma$ vertices}
\label{sec.ComplexVertices}

Below we consider diagrams containing $WW\gamma\gamma$, $WW\gamma$ vertices only because
their contributions are associated with logarithmic enhancement.
Let's consider the Feynman diagram with virtual photon which is emitted from the initial electron and absorbed
by the final electron (see Fig.~\ref{fig.e}).
The relevant contribution to the vertex function is
\eq{
\Delta V^e_\mu&=ie\frac{4\pi\alpha g^2}{2(16\pi^2)^2}V^e_\mu, \nn \\
V_\mu^e&=\int\frac{d^4 k}{i\pi^2}\frac{1}{(k^2-\lambda^2)(k^2-2p_1k)(k^2-2p_1'k)}
\int\frac{d^4 k_1}{i\pi^2}\frac{V_{\sigma\mu\eta}N^{\eta\sigma}}{k_1^2((k+k_1-p_1)^2-m_W^2)((k+k_1-p_1')^2-m_W^2)},
}
where $\lambda$ is the photon mass and
\eq{
V_{\sigma\mu\eta}&=g_{\sigma\mu}(2p_1-p_1'-k-k_1)_\eta+g_{\mu\sigma}(2p_1'-p_1-k-k_1)_\sigma+g_{\eta\sigma}(-p_1-p_1'+2(k+k_1))_\mu, \nn \\
N_{\eta\sigma}&=\bar{u}(p_1')\gamma_\lambda(\hat{p}_1'-\hat{k})\gamma_\eta\hat{k}_1\gamma_\sigma(\hat{p}_1-\hat{k})\gamma^\lambda\omega_-u(p_1).
}
Doing the similar treatment as it was done above one can integrate over loop momentum $k_1$, renormalize the amplitude of this subgraph and obtain
\eq{
V^e_\mu=\frac{3Q^2}{2m_W^2}\int\frac{d^4k}{i\pi^2}\left.\frac{\bar{u}(p_1')\gamma_\lambda(\hat{p}_1'-\hat{k})\gamma_\mu(\hat{p}_1-\hat{k})\gamma^\lambda\omega_-u(p_1)}
{(k^2-\lambda^2)(k^2-2p_1k)(k^2-2p_1'k)}\right|_{|k^2|\ll m_W^2}.
}
After integration over $k$ one gets:
\eq{
V^e_\mu=\frac{3Q^2}{2m_W^2}\int\limits_0^1 d x\int\limits_0^1 2y d y \brs{\ln\frac{m_W^2}{D_e}-\frac{Q^2}{D_e}(\bar{y}+y^2x\bar{x})},\qquad
D_e=y^2b^2+\lambda^2\bar{y}.
}
Further we use simple integrals 
\ga{
\int\limits_0^1\frac{2 y d y}{y^2b^2+\lambda^2\bar{y}}=\frac{1}{b^2}\ln\frac{b^2}{\lambda^2},\qquad
\int\limits_0^1\frac{Q^2}{b^2}d x=2\ln\frac{Q^2}{m^2},\nn\\
\int\limits_0^1\frac{Q^2x\bar{x}}{b^2}d x=1,\qquad
\int\limits_0^1\frac{Q^2}{b^2}\ln\frac{b^2}{m^2}\,d x=\ln^2\frac{Q^2}{m^2}-\frac{\pi^2}{3},\nn
}
and obtain
\eq{
\Delta V^e_\mu&=ie\frac{3Q^2}{2m_W^2}\bar{u}(p_1')\gamma_\mu\omega_-u(p_1)\frac{4\pi\alpha g^2}{2(16\pi^2)^2}I^e, \nn \\
I^e&=2+\frac{\pi^2}{3}+\ln\frac{m_W^2}{Q^2}-2\ln\frac{Q^2}{m^2}(\ln\frac{m^2}{\lambda^2}-2)-\ln^2\frac{Q^2}{m^2} = -2\ln\frac{Q^2}{m^2}\ln\frac{m^2}{\lambda^2}-40.388.
}

The diagrams in Figs.~\ref{fig.f},\ref{fig.g},\ref{fig.h} has a general enhancement factor which is associated with the collinear photon emission in vertex.
Let's demonstrate this in general. The common structure for all three diagrams
contains the emission of photon with momentum $k$ from initial electron. This leads to the following structure of the amplitude:
\eq{
V^{f,g,h}_\mu=\frac{e}{16\pi^2}\int\frac{d^4 k}{i\pi^2} \,\frac{\bar{u}(p_1')O^\lambda_\mu(\hat{p}_1-\hat{k}+m)\gamma_\lambda u(p_1)}{k^2(k^2-2p_1k)},
}
where $O^\lambda_\mu$ corresponds to the remaining part of Feynman diagram and different for each diagram.
We note that the dominant contribution to this integral comes from the integration region of small photon momentum (i.e. $|k^2| \ll m_W^2$) and thus
we can omit $k$ in the remaining part of vertex amplitude, containing the momenta of a $W$-boson.
Joining the denominators we have
\eq{
V^{f,g,h}_\mu&=\frac{e}{16\pi^2} \int\limits_0^1 d x \, 2 \bar{x} \, \bar{u}(p_1') p_{1\lambda}\left.O^\lambda_\mu\right|_{k \sim xp_1}u(p_1)\int\frac{d^4 k}{i\pi^2}\left.\frac{1}{\br{(k-p_1x)^2-m^2x^2}^2}\right|_{|k^2| \ll m_W^2}, \nn
}
and this approximately equals
\eq{
    V^{f,g,h}_\mu&\approx R \cdot  \bar{u}(p_1') {p_1}_\lambda \left.O^\lambda_\mu\right|_{k \sim xp_1} u(p_1),\qquad R \approx \frac{e}{16\pi^2}L,\qquad L=\ln\frac{m_W^2}{m^2}.
}

The diagram containing the $WW\gamma\gamma$ vertex (see Fig.~\ref{fig.f}) gives
\eq{
\Delta V^f_\mu&=2ie R\frac{4\pi\alpha g^2}{32\pi^2}S_{\mu\nu}^{~~\lambda\sigma}p^\nu_1\int\frac{d^4k_1}{i\pi^2}\frac{N^f_{\lambda\sigma}}{k_1^2 \br{k_1^2-2p_1k_1-m_W^2}\br{k_1^2-2p_1'k_1-m_W^2}}, \\
N^f_{\lambda\sigma}&=\bar{u}(p_1')\gamma_\sigma\hat{k}_1\gamma_\lambda\omega_-u(p_1), \qquad
S_{\mu\nu\lambda\sigma}=2g_{\mu\nu}g_{\lambda\sigma}-g_{\mu\lambda}g_{\nu\sigma}-g_{\mu\sigma}g_{\nu\lambda}. \nn
}
The loop momentum integral do not have ultraviolet as well as infrared divergences. Standard manipulations lead to
\eq{
\Delta V^f_\mu=2ie\frac{Q^2}{4m_W^2}\frac{4\pi\alpha g^2}{2(16\pi^2)^2}\, L\,\bar{u}(p_1')\gamma_\mu\omega_-u(p_1).
}

For the Feynman diagram with $WW\gamma$ vertex shown in Fig.~\ref{fig.g} we have
\eq{
\Delta V^g_\mu&=-2iR\frac{4\pi\alpha g^2}{32\pi^2} \int\frac{d^4 k_1}{i\pi^2}\frac{\bar{u}(p_1')\gamma_\eta\hat{k}_1\gamma_\lambda\omega_-u(p_1)V_1^{\lambda\nu\sigma}{V_{2}}_\mu^{~\eta\sigma}~p_{1\nu}}
{k_1^2 \br{k_1^2-2p_1k_1-m_W^2}^2 \br{k_1^2-2p_1'k_1-m_W^2}},
}
where vertices have the form
\eq{
V_1^{\lambda\nu\sigma}&=(2p_1-k_1)^\lambda g^{\nu\sigma}+(2k_1-p_1)^\nu g^{\sigma\lambda}+(-p_1-k_1)^\sigma g^{\lambda\nu}, \nn \\
{V_{2}}_\mu^{~\eta\sigma}&=(-p_1-p_1'+2k_1)_\mu g^{\eta\sigma}+(2p_1-p_1'-k_1)^\eta g^{\sigma}_\mu+(2p_1'-p_1-k_1)^\sigma g_\mu^\eta.
}
Retaining in the numerator the terms quadratic over loop momenta one gets
\eq{
\Delta V^g_\mu =-iR\frac{\alpha g^2}{2\pi}
\bar{u}(p_1')\gamma_\mu \omega_-u(p_1)\int\limits_0^1 x d x\int\limits_0^1 y^2 d y
\frac{Q^2}{D_g}\br{-\frac{13}{4}}, \qquad D_g \approx y m_W^2.
}
Finally we have for the contribution of Feynman diagram shown in Fig.~\ref{fig.g}:
\eq{
\Delta V^g_\mu=-ie\frac{Q^2}{m_W^2}\,L\,\frac{\alpha g^2}{32\pi^3}\frac{13}{36}\bar{u}(p_1')\gamma_\mu\omega_-u(p_1),
}
and the diagram shown in Fig.~\ref{fig.h} gives the similar result
\eq{
\Delta V^h_\mu=-ie\frac{Q^2}{m_W^2}\,L\,\frac{\alpha g^2}{64\pi^3}\frac{67}{36}\bar{u}(p_1')\gamma_\mu\omega_-u(p_1).
}

\section{Numerical contribution to the left-right asymmetry}
\label{sec.NumericalResults}

Collecting the result of considered two-loops contributions one can put the total result in the form
\eq{
    \Delta V_\mu^{a+b} + \Delta V_\mu^{{\rm MO}} + \Delta V_\mu^{e+f+g+h} = C^Z K_\mu^Z + C^W K^W_\mu,
}
where
\eq{
K^Z_\mu&=ie\frac{Q^2}{m_Z^2}(1\pm a)^2\frac{\alpha g^2}{(4c_W)^2256 \pi^3}\bar{u}(p_1')\gamma_\mu u(p_1), \\
K^W_\mu&=ie\frac{Q^2}{m_W^2}\frac{\alpha g^2}{256 \pi^3}\bar{u}(p_1')\gamma_\mu\omega_- u(p_1),
}
and the coefficients look like
\eq{
C^Z&=-8(c_1^b+c_2^b)+32 c_3^Z =70.5285, \nn \\
C^W&=-4(c_1^a+c_2^a)+8 c_3^W+3I^{e,fin}+ L\br{1-\frac{26}{9}-\frac{67}{9}}=-312.382.
}

Let's note by index $C$ the contributions investigated here, i.e. $C=a,b,...,h$.
As specific corrections to observable parity-violating asymmetry induced by contribution $C$
we choose the contribution to the asymmetry $(\Delta A)_C$
and the relative corrections $D_A^C$:
\eq{
(\Delta A)_C=\frac{|\M_C^{----}|^2-|\M_C^{++++}|^2}{\sum |\M_0^{\lambda}|^2},
\qquad
D_A^C = \frac{(\Delta A)_C}{A^{0}} = \frac{|\M_C^{----}|^2-|\M_C^{++++}|^2}{|\M_0^{----}|^2-|\M_0^{++++}|^2}.
}
The physical effect of radiative effects from contribution $C$ to observable asymmetry
is determined by the relative correction  (see \cite{Aleksejevs:2015dba} for more details):
\eq{
\delta_A^{C}
= \frac{A_{}^{C}-A_{}^0}{A_{}^0} = \frac{ D_A^C -\delta^C }{1+\delta^C},
\label{addit}
}
where the relative correction to unpolarized cross section is $\delta^C= {\sigma^{C}_{00}}/{\sigma^0_{00}}$.
For two-loop effects (where $\delta^C$ is small)
the approximate equation for relative correction to asymmetry takes place: $\delta^{C}_A  \approx  D_A^C$.

Contributions to asymmetry of $Z$ and $W$ types are
\eq{
(\Delta A)_Z=-16 a C^Z\frac{Q^2}{m_Z^2}\frac{\alpha g^2\pi}{(4c_W)^2(16\pi^2)^2},\qquad
(\Delta A)_W=4C^W\frac{Q^2}{m_W^2}\frac{\alpha g^2\pi}{(16\pi^2)^2},
}
which give the relevant numerical values:
\eq{
(\Delta A)_Z = -2.5410 \cdot 10^{-12},
\qquad
(\Delta A)_W = -3.1983 \cdot 10^{-10}.
}
Taking into account that in CR of MOLLER the Born asymmetry $A_{}^0$ = 94.97 ppb
the numbers for relative corrections $D_A^C$ are
\eq{
D_A^Z = -0.0000267,
\qquad
D_A^W = -0.0033677.
}
We can see that effects have the same negative sign, first is rather small,
but the second one is at the edge of region of planned one per cent experimental error for MOLLER
and thus will be important for future analysis of MOLLER experimental results.

\section{Acknowledgements}

The work of A.G.A. and S.G.B. has been supported by the Natural Science and Engineering Research Council of Canada (NSERC).
Yu.M.B. acknowledges support of Heisenberg-Landau Program grant No. HLP-2015-15.
V.A.Z. is grateful to the financial support of Belarus program "Convergence" (no. 20141163)
and thanks JINR for hospitality in 2014--2015.


\end{document}